 \documentstyle[preprint,aps]{revtex}

\def\hef{$^4\!$He }
\def\rr{ {\bf r} }
\def\rp{ {\bf r'} }

\begin{document}

\draft
\preprint{UTF 326/94  }

\title{ Dispersion of Ripplons in Superfluid $^4$He }

\author{ A. Lastri, F. Dalfovo, L. Pitaevskii\cite{Moscow},
         and S. Stringari }

\address{  Dipartimento di Fisica, Universit\`a di Trento, \\
           I-38050 Povo, Italy }

\date{ April 15, 1994 }
\maketitle

\begin{abstract}
A detailed study of the dispersion law of surface excitations in
liquid \hef at zero temperature is presented,  with special emphasis
to the short wave length  region. The hybridization mechanism between
surface and bulk modes is discussed on a general basis, investigating
the scattering of slow rotons from the surface. An accurate density
functional, accounting for backflow effects,  is then used to determine
the  dispersion of both bulk and surface  excitations. The numerical
results are close to  the experimental data obtained on thick films and
explicitly reveal the occurrence of important hybridization effects
between ripplons and  rotons.
\end{abstract}


%
%

\section{ Introduction }
\label{intro}

Superfluid helium is a unique system where undamped excitations can
propagate also in the short wave length regime. This behaviour
characterizes  the dynamics of the bulk as well as of the surface,
and is at the origin of several interesting phenomena. Hybridization
between ripplons and rotons  and quantum evaporation are important
examples of such phenomena where the surface of helium is explicitly
involved. These peculiar properties of liquid helium at low temperature
have been  the object of several experimental investigations in the
last years.   First results for the dispersion of short wave length
ripplons have recently become available via neutron scattering
experiments  on  helium films \cite{Lau90}. Quantum evaporation, i.e.,
the emission of an atom produced by a roton  propagating  balistically
in the medium and impinging on the surface, has been also systematically
investigated (see for example Ref.\ \cite{Wya92}).

The above phenomena have received attention also from
a teoretical point of view. However, due to intrinsic difficulties
associated  with a reliable description of the statics and dynamics
of strongly interacting inhomogeneous systems, the predictions
have been so far only  semi-quantitative.
The main reason lies in the fact that both microscopic approaches,
based on correlated basis functions, and phenomelogical methods,
mainly based on density functional theory, have been used in
the framework of the so called Feynman approximation, or equivalent
schemes. This approximation significantly overestimates the roton energy
and this  clearly rules out the possibility of a quantitative description of
quantum evaporation as well as of the dynamics of energetic ripplons,
since in both cases the relevant energies are of the order of the
roton gap.

In order to obtain a better description it is necessary
to include backflow effects in the theory.    Here, following
the spirit of  density functional theory, we account
for such effects through a new phenomenological current dependent
interaction  \cite{trento-orsay}. In the present work we
use the new functional to provide a quantitative description of
the ripplon dispersion, with special emphasis to the region near
the roton threshold. The study of quantum evaporation will be
the object of a future work.

The work is organized in two parts:
in Sec.\ \ref{hybri}  we discuss in details the mechanism of the
hybridization between ripplons and rotons. In particular we
show that this mechanism is governed by the properties of
reflection of rotons from the free surface.
In Sec.\ \ref{tddf} we present numerical results for
the ripplon dispersion
obtained by solving the equations of time dependent density
functional theory with the inclusion of backflow terms.

\section{Hybridization between ripplons and rotons}
\label{hybri}

The hybridization between ripplons and rotons consists of a
coupling between surface and bulk excitations that takes place when
the energy of ripplons is close to the one of rotons.

The physical origin of such a phenomenon is clear. Let us consider
a surface excitation with energy larger than the roton gap $\Delta$. This
surface mode can naturally decay into a roton if its momentum parallel
to the surface is lower than the roton momentum. Indeed this process
is possible because only the  momentum parallel to the surface is
conserved, while the orthogonal momentum can be transferred to the
surface.

This decay mechanism provides a natural threshold for the energy of
ripplons, given by the roton gap $\Delta$.  The occurrence of this
threshold has been
confirmed by recent experimental data with neutron scattering \cite{Lau90}
and is clearly expected to influence the dispersion of energetic
ripplons. Such a
phenomenon was first explicitly explored in the case of helium
in Ref.\ \cite{Pit92} using the formalism of Green's functions.
A similar phenomenon was discussed in Ref.\ \cite{Mar73} for the case of
dieletrics where it results in a damping of surface waves.

The problem of the interaction between ripplons and bulk excitations in
superfluid $^4$He was  investigated several  years ago by Edwards et al.
\cite{Edw74,Edw78}. They wrote the solution for the velocity potential
of the ripplon in the form
\begin{equation}
\phi = A e^{i(k_xx + k_zz -\omega t)}
\label{phi}
\end{equation}
where $x$ and  $z$ are parallel and orthogonal to the surface, respectively;
the  wave vector $k_z$ is a function of $\omega$ and $k_x$, defined  by the
equation
\begin{equation}
\omega = \epsilon( (k_x^2 + k_z^2)^{1/2})
\label{omega}
\end{equation}
where $\epsilon(k)$ is the dispersion law for bulk excitations.
The authors of Ref.\ \cite{Edw74,Edw78} took  only imaginary solutions
of  Eq.\ (\ref{omega}) for $k_z$.
This corresponds to consider purely damped waves in the
$z$-direction with damping length $|k_z|^{-1}$. The solution is then
given by $k_z = i \sqrt{k^2_x - k^2(\omega)}$ where $k(\omega)$
is the momentum of bulk excitations with energy $\omega$. Of course this
solution  exists only for $k_x > k(\omega)$. However this  is not
the only possibility, since Eq.\ (\ref{omega}) can have also complex
solutions for $k_z$. In particular, when $ \omega$ is
below and close to the roton gap $\Delta$, one can use the well
known approximation \cite{note1}
\begin{equation}
\epsilon = \Delta + {( k - p_0)^2 \over 2 \mu}
\label{epsilon}
\end{equation}
to the roton dispersion, yielding the following solution for $k_z$:
\begin{equation}
 k_z \simeq p_1 + i {p_0 \over p_1} \sqrt{2 \mu
(\Delta - \omega)}
\label{kz1}
\end{equation}
with $p_1 = \sqrt{p_0^2- k^2_x}$.
This solution corresponds to a penetration depth
\begin{equation}
\delta =  {p_1 \over p_0 \sqrt{2 \mu  (\Delta - \omega)}}
\label{delta}
\end{equation}
that becomes larger and larger as $ \omega$ approaches $\Delta$ at
fixed $k_x$.

To investigate the behaviour of the ripplon dispersion near the
threshold $\Delta$ one must learn more about the dynamics of slow
rotons. For this we take the familiar roton Hamiltonian
\begin{equation}
H = \Delta + { (k^2 -p_0^2)^2 \over 8 \mu p_0^2 } \ \ \ ,
\label{h1}
\end{equation}
yielding dispersion (\ref{epsilon}) for $k$ close to $p_0$.
At given $k_x$ Eq.~(\ref{h1}) can be written in the form
\begin{equation}
H= \Delta + { (k_z^2 - p_1^2)^2 \over 8 \mu p_0^2 } \ \ \ ,
\label{h2}
\end{equation}
with $p_1$ defined above.  For the description of the threshold
behaviour of ripplons only rotons with energy close to $\Delta$ are
important. In this case the roton wave function can be written as
\begin{equation}
\Psi_r (z) = f_1(z) \exp [-i p_1 z] +
f_2(z) \exp [i p_1 z] \ \ \ ,
\label{psir1}
\end{equation}
where $f_1(z)$ and $f_2(z)$ are slowly varying function of $z$.
By neglecting higher derivatives of  $f_i$  in
the Schr\"odinger equation one finds that the functions $f_i$ are
solutions of
\begin{equation}
H_f f_i = \omega f_i \ \ \ \ , \ \ i=1,2
\label{hffi}
\end{equation}
with
\begin{equation}
H_f = \Delta - { 1 \over 2 m^*} {d^2 \over dz^2 }
\label{hf}
\end{equation}
and $m^*=\mu p_0^2/p_1^2$ \cite{note2}.

In the same scheme the surface mode, with energy close to $\Delta$,
appears as a surface bound state of rotons, whose wave function can
be written as
\begin{equation}
\Psi_r(z)=2A \exp[-\kappa z] \cos(p_1 z)
\label{psirrip}
\end{equation}
where
\begin{equation}
\kappa= [2 m^* (\Delta -\omega)]^{1\over2} \ \ ,
\label{kappa}
\end{equation}
and we have made an explicit choice for the phase. Equation (\ref{psirrip})
corresponds to taking
\begin{equation}
f_1(z) =f_2(z) = A \exp[-\kappa z] \ \ \ .
\label{f1f2}
\end{equation}
Clearly both the scattering and the bound solutions (\ref{psir1}) and
(\ref{psirrip}) hold only far from the surface.
The quantity $\kappa$ is a  function of $k_x$, which can
be found, in principle,  by solving the proper Schr\"odinger equation
in the surface  region.  Close to the surface, where $|z|$ is of  the
order of the surface  thickness $a$, the Hamiltonian (\ref{hf}) has
to be changed   in order to account for the interaction of rotons with
the surface.  However for slow rotons the solutions $f_i$ of the
Schr\"odinger equation vary slowly and the role of the surface
can be safely replaced  by the imposition of suitable boundary
conditions at $z=0$. The general form of these boundary conditions
is
\begin{mathletters}
\label{sandro2}
\begin{equation}
{df_1 \over dz } \bigg\vert_{z=0} = - \alpha_{11}  f_1(0)
- \alpha_{12} f_2(0)\
\label{s1:1}
\end{equation}
\begin{equation}
{df_2 \over dz } \bigg\vert_{z=0} = - \alpha_{21} f_1(0)
-\alpha_{22} f_2(0) .
\label{s1:2}
\end{equation}
\end{mathletters}
The matrix $\alpha_{ij}$ must be hermitian in order to ensure real values
for $\kappa$ in the case of bound states. Furthermore, due to time reversal
invariance, one must have $\alpha_{11}=\alpha_{22}$ and the phases of the
functions $f_1(z)$ and $f_2(z)$ can be always chosen such that
$\alpha_{12}=\alpha_{21}^*$ is real. The quantities $\alpha_{11}$ and
$\alpha_{12}$, fixing the boundary conditions at $z=0$, are
entirely determined  by the properties of the surface.

Let us first apply Eq. (\ref{sandro2}) to the surface mode for
$\omega< \Delta$. In this case the functions $f_i$ are given by
expression (\ref{f1f2}) so that Eq. (\ref{sandro2}) yields
\begin{equation}
\kappa = \sqrt{{2  m^* (\Delta -\omega) }} = \alpha(\omega,k_x)
\ \ \ ,
\label{kappaoverm1}
\end{equation}
where $\alpha(\omega,k_x)=\alpha_{11} + \alpha_{12}$.
This equation has a solution describing a surface mode if $\alpha > 0$.
This scheme provides a natural description of the dispersion near
the threshold. Let us suppose, for example, that
$\alpha(\Delta,k_x)=0$ at some point $k_x=k_t<p_0$. Near this point the
function $\alpha$ can be expanded as $\alpha(\Delta,k_x) \simeq
\beta (k_x -k_t)$. Thus the dispersion curve near the threshold  has a
parabolic form:
\begin{equation}
\Delta - \omega =  { \beta^2 \over 2 m^*} (k_x - k_t)^2 \ \ \ ,
\label{delta-omega}
\end{equation}
approaching the point $k_t$ with a horizontal tangent.

When $\omega > \Delta$ the boundary conditions (\ref{sandro2})
describe the reflection of rotons at the
surface. In this case the functions $f_i$ take the form
\begin{equation}
f_1 = \exp[-i k_z z] + R_1(\omega) \exp[ik_z z]
\label{f1}
\end{equation}
\begin{equation}
f_2= R_2(\omega) \exp[ik_z z]  \ \ \ .
\label{f2}
\end{equation}
By definition the coefficients $R_1$ and $R_2$ are the amplitude for
change-mode reflection and normal reflection, respectively.
After imposing the boundary condition (\ref{sandro2})
in  the threshold region $\omega \simeq \Delta$ and near $k_t$
where $\alpha\ll \alpha_{11},\alpha_{12}$, we find
the following results for the amplitudes $R_1$ and $R_2$:
\begin{equation}
R_1  = - { \alpha  \over ik_z + \alpha }
\label{r1omega}
\end{equation}
and
\begin{equation}
R_2  = { ik_z \over ik_z + \alpha}
\ \ \ .
\label{r2omega}
\end{equation}
Note that if $k_z \ll \alpha$ one has
\begin{equation}
R_1  \simeq -1 + i { k_z \over \alpha  } \ \ \ .
\label{ralmost1}
\end{equation}
This equation shows that when $\omega$ approaches $\Delta$, and
consequently $k_z \to 0$, the change-mode reflection amplitude
approaches the  value $-1$, while the normal reflection amplitude
tends to zero \cite{note3}. In this limit the
solution (\ref{psir1}) of the Schr\"odinger equation, far
from the surface, takes  the form
\begin{equation}
\Psi_r (z) = \exp [-i (p_1 +k_z)z ] - \exp [ -i (p_1 - k_z) z ]
\label{psir2}
\end{equation}
and corresponds to the
reflection of a roton R$^+$, with momentum $(k_x,0,p_1+k_z)$, into a roton
R$^-$, with momentum $(k_x,0,p_1-k_z)$ (see Fig.~\ref{change-mode}).
Notice that the roton R$^-$ has negative velocity and,consequently, the roton
scatters backwards not only in the $z$ direction, but also in the
$x$ one. The occurrence of the change-mode scattering, peculiar of rotons,
will be explicitly confirmed in Sec.\ \ref{tddf} through
the numerical solution of the equations of the time dependent
density functional theory.

The theory developed above is very general. It is reasonable to assume
however the interaction between ripplons and rotons in \hef, producing
the hybridization, to be weak. In this case one can proceed further,
looking for a theory which describes the crossover region between an
unperturbed ripplon dispersion, entirely characterized by surface properties
in the absence of interaction with bulk modes, and the dispersion near
the threshold, dominated by the hybridization mechanism. Under the
assumption of weak interaction  one expects (see the Appendix) that
the function $\alpha$ has the form
\begin{equation}
\alpha(\omega,k_x) = b(k_x) + {g^2 \over \epsilon_0(k_x)-\omega}
\label{alphabg}
\end{equation}
when $\omega$ is close to $\Delta$. In Eq. (\ref{alphabg}) $\epsilon_0
(k_x)$ is the dispersion law of unperturbed ripplons and $g$ is a coupling
constant. By using Eq. (\ref{alphabg}) in Eq. (\ref{kappaoverm1}) the
dispersion of the surface modes becomes
\begin{equation}
\sqrt{2 m^* (\Delta - \omega) } \  = \
b(k_x) + {g^2 \over \epsilon_0(k_x)-\omega} \ \ \ .
\label{sqrt}
\end{equation}
Note that the coefficient $g^2$ must be  positive to ensure a
stable solution of Eq. (\ref{sqrt})  when the unperturbed
ripplons have a damping. Different scenarios  can now take place depending
on the value of the coefficient  $b$:

\begin{itemize}
\item 1)  $b<0$.  In this case the spectrum reaches the
$\omega=\Delta$ threshold at a point
$k_x=k_t$, as assumed in the derivation of result (\ref{delta-omega}). The
value of $k_t$ is fixed by the equation
\begin{equation}
{g^2 \over \epsilon_0(k_t) - \Delta} = - b(k_t) \ \ \ ,
\label{kt}
\end{equation}
and the coefficient $\beta$ in Eq. (\ref{delta-omega}) can be
directly calculated from Eq. (\ref{sqrt}).
Clearly the point $k_t$ lies at the right of the unperturbed ripplon
branch.

\item 2)  $b>0$.
Equation (\ref{sqrt}) has now a solution also for $g=0$. A surface
branch exists independently of the ripplon dispersion. This branch
corresponds to a bound state of rotons near the surface. This
scenario describes a hybridization between two surface modes,
ripplons and surface rotons respectively,
near the threshold for bulk modes.

\item 3) If the coefficient $b$ satisfies the
condition  $|b| \ll (m^*)^{1/3} g^{2/3} $, then its role can be neglegted
in Eq. (\ref{sqrt}). One obtains that the ripplon dispersion approaches
asymptotically the roton threshold with the law
\begin{equation}
\Delta - \omega \propto  { g^4 \over ( \epsilon_0(k_x) -
\Delta)^2 } \ \ \ .
\label{oldscenario}
\end{equation}
This hybridization mechanism was previously discussed in Ref. \cite{Pit92}.
Also note that a transition from $b$ small and positive to $b$ small
and negative corresponds to a transition from a bound surface state to
a "virtual level" (see \cite{Landau} \S 133).
\end{itemize}

\bigskip

In the next Section we will carry out a full calculation of the
ripplon dispersion using a density functional approach. The numerical
results will be shown to follow the behaviour predicted by the first scenario
(Eqs. (\ref{delta-omega}) and  (\ref{kt})).

\section{ Predictions of density functional theory  }
\label{tddf}

In this section we present the results of a calculation of the dynamics
of inhomogeneous liquid helium obtained using time dependent density
functional  theory.   In the density functional  approach the internal
energy is assumed to
be a functional of the  one-body densities in the form
\begin{equation}
E \  = \  \int \! d\rr \  {\cal E} [\Psi^* \Psi]
\label{energy}
\end{equation}
where the wave function $\Psi$,  for  a Bose system at zero temperature,
is written as
\begin{equation}
\Psi (\rr,t) = \Phi (\rr,t) \exp \left( i S(\rr,t) \right)
\ \ \ .
\label{psi}
\end{equation}
The real function $\Phi$ is related to the particle density by
$\rho = \Phi^2$, while the phase $S$ is related to the velocity
$\bf v$ of the fluid by
\begin{equation}
{\bf j} = \rho {\bf v} = {i  \over 2 m} (\Psi \nabla \Psi^*
- \Psi^* \nabla \Psi) = {\rho \over m} \nabla S \ \ \ ,
\label{j}
\end{equation}
where $\bf j$ is the current density, and $m$ is the mass of the \hef
atoms.

The formalism of time dependent density functional has been already
applied to describe the surface excitations of superfluid helium
\cite{Kro86,Kro92,Ji86,Pri93}. In Ref.\ \cite{Kro92} the effective
Hamiltonian has been derived microscopically starting from the
interatomic potential in the framework of the hypernetted chain
scheme. Vice-versa the functionals employed in  Refs.\ \cite{Ji86,Pri93}
are of phenomenological type.

The equations of motion of time dependent density functional theory
can be derived starting from the least action principle
\begin{equation}
\delta \int_{t_1}^{t_2} dt \int d\rr \left[ {\cal E} [\Psi^* \Psi]
- \Psi^* i  {\partial \Psi \over \partial t}
\right] \  = \ 0 \ \ \ ,
\label{leastaction}
\end{equation}
making variations with respect to $\Phi$ and $S$ \cite{Pri93}, or,
equivalently,  with respect to $\Psi^*$. In the latter case one finds a
Schr\"odinger-like equation of the form
\begin{equation}
\tilde{H}  \Psi = i {\partial \over \partial t } \Psi \ \ \ ,
\label{htilde}
\end{equation}
where $\tilde{H} = \delta {\cal E} / \delta \Psi^*$ is an effective
Hamiltonian.  If one looks for linearized solutions
\begin{equation}
\Psi ({\bf r},t) = \Psi_0({\bf r},t) + \delta \Psi({\bf r},t)
\label{deltapsi}
\end{equation}
the Hamiltonian  $\tilde{ H}$ takes the form
\begin{equation}
\tilde{H} = \tilde{H}_0 + \delta \tilde{H} \ \ \ .
\label{deltah}
\end{equation}
The static Hamiltonian $\tilde{H}_0$ is fixed by the equilibrium
state  $\Psi_0({\bf r},t) = e^{-i\mu_4 t}\sqrt{\rho(z)}$ ($\mu_4$
is the chemical potential) through the equation  $\tilde{H}_0 \Psi_0=
\mu_4 \Psi_0$. The term $\delta \tilde{H}$ is linear in $\delta
\Psi$ and accounts for changes in the Hamiltonian induced by the collective
motion of the system. Since $\tilde{H}$ depends explicitly
on the wave function $\Psi$, the Schr\"odinger
equation (\ref{htilde}) has to be solved using a self-consistent
procedure,  even in the linear limit considered in the present work.
The formalism then concides with the one of the Random  Phase Approximation
(RPA) for Bose systems. Of course this theory, which is basically
a mean field theory, accounts for collective and single
particle excitations, but  cannot account for multiphonons
excitations.

In the bulk  $\Psi_0$ is constant and $\delta \Psi$ can be expanded
in plane waves, corresponding to the propagation of the phonon-roton mode.
In  the presence of a free surface, orthogonal to the $z$-coordinate,
one can write
$\Psi$ as
\begin{equation}
\Psi (x,z,t) = \Psi_0(z,t) +
            e^{-i \mu_4 t} \left(
            \Phi_1(z) e^{-i(\omega t - k_x  x)} +
            \Phi_2(z) e^{ i(\omega t - k_x  x)}
                        \right)  \ \ \ ,
\label{psixzt}
\end{equation}
where the structure of $\Phi_1$ and $\Phi_2$ allows to distinguish between
bulk and surface modes. The linear variation of the density (transition
density),  associated with the solution (\ref{psixzt}), is given by
\begin{equation}
\delta \rho = 2 |\Psi_0 (z)| \Phi^+ (z) \cos(\omega t - k_x x)
\label{drho}
\end{equation}
with $\Phi^+ = \Phi_1 + \Phi_2$.

{}From the equation of motion (\ref{htilde}) we obtain linear
equations for the functions $\Phi_1 (z)$ and $\Phi_2 (z)$,
which, without any loss of generality, can be chosen real.
These  functions  are expanded on a  basis of eigenstates of the static
one-body Hamiltonian $\tilde{H}_0$ in order to get a matrix equation
which is  solved numerically by direct diagonalization.

In order to write the RPA equations one needs the explicit form of the
functional $\cal E$. The functionals used so far
in liquid helium have the form \cite{Kro86,Ji86,Pri93,Str87}
\begin{equation}
{\cal E}_0 \  =  \   {1 \over 2m} | \nabla \Psi |^2
+ {\cal V} [\rho] \ \ \ ,
\label{cale0}
\end{equation}
where ${\cal V } [\rho]$ is a velocity independent functional. An
accurate phenomenological functional of this form \cite{Dup90}
(hereafter named Orsay-Paris functional)  has been used,
for instance, in Ref. \cite{Pri93} to study the dispersion of
collective modes in helium films.  The same functional has proven
to be quite reliable in the study of the equilibrium configuration of
several helium systems, such as the free surface \cite{Dup90}, films
\cite{Che91}, droplets \cite{Dal94}, and vortices \cite{Dal92}.

With  functionals of the form (\ref{cale0}) the dispersion law
predicted by Eq. (\ref{htilde}) has the form
\begin{equation}
\omega^2 = {k^2 \over m |\chi(k)|} \ \ \ ,
\label{omfey}
\end{equation}
where $\chi$ is the static response function. The quantity $\chi(k)$ is
a key ingredient for calculations with density functional theory
at zero temperature. A major advantage of liquid \hef is that $\chi(k)$
is well known experimentally \cite{Cow71}.

It is worth noticing that the dispersion
law (\ref{omfey}) does not reproduce correctly the experimental phonon-roton
dispersion, especially in the roton region. This is
due to the fact that functional (\ref{cale0}) ignores backflow effects.
To understand better this point it is useful to rewrite
Eq.\ (\ref{omfey}) as the ratio between the energy weighted
and the inverse energy weighted  moments of the dynamic structure
function: $\omega^2= m_1(k)/m_{-1}(k)$,
where
\begin{equation}
m_1(k) = \int d\omega \ S(k,\omega) \ \omega = {k^2 \over 2m}
\label{fsumrule}
\end{equation}
is the f-sum rule, and
\begin{equation}
m_{-1}(k)= \int d\omega \ S(k,\omega) \ \omega^{-1} =-  {1\over 2} \chi(k)
\end{equation}
is the compressibility sum rule. The analysis of the spectra
of neutron scattering experiments \cite{Cow71} shows that, due to
the $\omega^{-1}$  factor in the integrand, the collective
phonon-roton mode almost exhausts the compressibility sum rule for
all wave lengths up to about $2.2$ \AA$^{-1}$.
On the contrary, it gives only a fraction
($\simeq  1/3$) of the energy weighted sum rule, the remaining part
being exhausted by high energy multiphonon excitations. This explains
why functionals of the form  (\ref{cale0}), yielding the
dispersion law (\ref{omfey}),  significantly overestimate
the roton energy.

To overcome this difficulty we add a current-current interaction term
to the functional (\ref{cale0}):
\begin{equation}
{\cal E} = {\cal E}_0 - {m \over 4} \int \! d\rp \
V_J (|\rr-\rp|) \ \rho(\rr) \rho(\rp) ({\bf v}(\rr) -{\bf v}(\rp))^2
\ \ \ .
\label{cale}
\end{equation}
The new term is Galilean invariant and depends on local changes in
the fluid  velocity field. The new dispersion of bulk excitations, obtained
by solving Eq.\ (\ref{htilde}), takes the form
\begin{equation}
\omega^2 = {k^2 \over m |\chi(k)| } \left[ 1- \rho
( V_J(k) - V_J(0) ) \right] \ \ \ ,
\label{hbaromega}
\end{equation}
where $V_J(k)$ is the Fourier transform of the current interaction
$V_J(r)$  of Eq.\ (\ref{cale}). Notice that the new term in Eq. (\ref{cale})
does not modify the static response function $\chi(k)$, which is
completely fixed by ${\cal E}_0$.  Notice also that the dispersion
law (\ref{hbaromega})  is not affected by the new term in
the $k \to 0$  hydrodynamic regime,  where it coincides with
the traditional phonon law $\omega = k/\sqrt{m |\chi(0)|}$.
This is an important feature ensured by Galilean invariance.

Equation (\ref{hbaromega}) explicitly shows that
the inclusion of the current  term in the density functional
can provide a reduction of the  phonon-roton
contribution to the moment $m_1$,  thereby improving the
description  of the dispersion law.
The fact that the new functional no longer satisfies the
f-sum rule (\ref{fsumrule}) points out in a clear way that
the time dependent density functional theory, or
equivalently the RPA,  does not account for
multiphonon excitations  which provide the remaining fraction of
the f-sum rule. A similar separation between collective
and multiphonon excitations, in the context of linear response
theory, has been developed by Pines and collaborators (see for
example Ref. \cite{Pin83}).

Let us briefly  discuss the criteria used to choose the new
functional, whose explicit form and parametrization is discussed
in Ref. \cite{trento-orsay}. The  static part $\cal V$ has a form
similar to the one of the  Orsay-Paris  functional \cite{Dup90}.
It contains a two-body
Lennard-Jones  interaction and a phenomenological density dependent
term accounting  for short range  correlations. The density dependent
term of the Orsay-Paris functional has been changed in order to
reproduce more accurately the experimental value of the static
response function $\chi(k)$ in the roton region.
The current-current interaction $V_J$ is then chosen
to reproduce phenomenologically the phonon-roton dispersion in bulk liquid
\cite{Sti90} through Eq. (\ref{hbaromega}). Its form is shown
in Fig.~\ref{vj}.

After fixing the density functional to reproduce the bulk spectrum, one can
solve the equations of motion in systems with non uniform density. We
will focus here on the problem of the free surface.
We choose, for numerical convenience,
a slab geometry (liquid between two parallel surfaces) and look for
solutions of the form (\ref{psixzt}). All the  states are confined in a
box larger  than the slab thickness, and, consequently, even the continuum
of states  outside the slab is discretized. We take  typically a basis
of more than 50 eigenstates of ${\tilde H}_0$, to  cover all the relevant
part of the spectrum.

The main features of the spectrum, extrapolated to the semi-infinite
system,  are shown in Fig.~\ref{schematic}. The dispersion of
bulk phonon-roton modes is the upper solid line; the agreement
with the  experimental data follows from the choice made
for $V_J$. The surface mode corresponds to the lowest solid line.
An important feature emerging from the present calculation
is the  deviation of the ripplon dispersion from the
hydrodynamic law. This is  in agreement with the
experimental data for surface excitations in helium
films  \cite{Lau90}. The horizontal line
corresponds to the threshold for rotons having energy close to
$\Delta=8.65$ K and propagating at different angles. The
dispersion of the surface mode reaches the roton threshold at about
$1.15$ \AA$^{-1}$. The threshold $ \omega = |\mu_4| +  k_x^2 /2m$
for creating  excited states outside the liquid, at zero temperature,
is also  shown. The  value at $k_x=0$  coincides with the experimental
chemical  potential,  $|\mu_4|=7.15$ K.

To  appreciate the effect of the backflow term included in the
functional, the
spectrum obtained  without backflow ($V_J \equiv 0$) is plotted
in Fig.~\ref{feynman}. The roton minimum  is pushed up
significantly, from $8.65$ K to about $14$ K, as follows
from Eq. (\ref{hbaromega}), while the dispersion of the surface
mode is less affected by backflow. As a consequence the branch
of the surface mode reaches $\Delta$  near the roton minimum.
The spectrum in Fig.~\ref{feynman} is similar to the one of
previous  calculations on  slabs \cite{Ger92,Pri93}  and films
\cite{Pri93,Kro92}. We note however that, even  without backflow,
density functional theory, once the value of $\chi(k)$ is properly
accounted for,  gives better predictions than the original Feynman
approximation. The latter, in fact, gives the roton minimum at
about $20$ K. The reason is that the phonon-roton dispersion
in Feynman approximation is fixed by the ratio $\omega=m_1(k)/m_0(k)$
between the energy weighted and non energy weighted moments of the
dynamic structure function. This ratio should be compared with
prediction (\ref{omfey}) of  density functional theory without backflow.
The moment $m_0$ is more affected  than the inverse energy
weighted moment $m_{-1}=-(1/2) \chi(k)$ by multiphonon
excitations. It then follows that  the  density functional approach
gives a significantly lower value for the energy  of the
collective mode ($14$ K at the roton minimum).

Due to the finite thickness of the slab, with parallel
surface at $z=\pm L$, one can distinguish  between odd and even states.
In Fig.~\ref{allstates} we show a  typical spectrum with the excited
states of a slab of thickness $2L=50$ \AA.  For simplicity, only  even  states
are shown in the figure.  Each state, at given energy and parallel wave
vector, is a stationary  superposition of bulk excitations,
surface excitations and free atoms, with the proper matching conditions.
For instance, the almost orizontal lines are mainly bulk rotons reflecting
at the slab surfaces; the gap between these states is fixed by the
slab thickness. The oscillations of the roton lines, below the maxon
region,  are peculiar of the slab
geometry; they originate from a periodic splitting of
even and odd states. The period of such oscillations is predicted
by the theory of roton scattering developed in Sect. \ref{hybri} to
be $\Delta k_x \simeq p_1 \pi /(2Lk_x)$.
The energy of the surface mode is instead practically independent of $L$.
Above the ripplon mode there is also an excited surface mode, whose
dispersion reaches the roton threshold at about $k_x=0.7$ \AA$^{-1}$.

The solution of the RPA equations also allows to compute  the dynamic
structure function,  defined by
\begin{equation}
S({\bf k},\omega) = \sum_n |\langle n| \rho_k^\dagger |0\rangle |^2
\delta(\omega-\omega_{n0})
\label{sqomega}
\end{equation}
where $\rho_k^\dagger=\sum_j \exp[i{\bf k}\cdot {\bf r}_j]$ is
the usual density  operator, and ${\bf k} \equiv (k_x,0,k_z)$.
An example is given in Fig.~\ref{strength} where we show the
results for $k_x=0.8$ \AA$^{-1}$ and $k_z=0$ in a slab $50$ \AA\
thick. The ripplon corresponds to the peak at lowest energy.
The other peaks are excited surface modes, as the one at
about $9$ K, and bulk modes, the most important
one being the high  energy phonon  at about $13$ K.
The figures reveals that, even for a rather thick slab, the
contribution  of the surface mode to the total strength is significant.
It is also interesting to explore the dependence of the relative
strength on the scattering angle, i.e. as a function of $k_z$
for a fixed $k_x$. In Fig.~\ref{angle} the relative contribution
of the surface mode to the total strength is shown for $k_x=0.8$
\AA$^{-1}$. One notes that the most favorable condition for exciting
ripplons is to induce momentum transfer parallel to the surface
($k_z=0$).

In Fig.~\ref{threeexamples}  the  quantity $\Phi^+$, characterizing
the transition density  (see Eq. (\ref{drho})), is plotted in three
cases: a surface mode (a),
free  atoms coupled  to high energy phonons (b),  free atoms
coupled to rotons (c).
In Fig.~\ref{gap}  we show in detail what happens to the surface mode
when its energy approaches the energy  of the lowest roton mode.
The surface mode transforms  continously in a bulk mode, and vice-versa.
The gap between the two levels tends to vanish when the slab thickness
increases. The resulting  hybridization mechanism then agrees with the
predictions of the scenario (1) discussed in Section \ref{hybri}
(see Eqs. (\ref{delta-omega}) and (\ref{kt})), with $k_t \simeq 1.15$
\AA$^{-1}$.

In Sec.\ \ref{hybri} we have shown that rotons close to $\Delta$,
according to Eqs. (\ref{r2omega}) and (\ref{ralmost1}), are totally
reflected with a change-mode reflection, as in Fig. \ref{change-mode}.
The roton wave function
for energy close to $\Delta$ is explicitly  written in Eq. (\ref{psir2}).
In the slab geometry one has to account for
$R^+$ and $R^-$ rotons propagating in both directions and reflecting
on two surfaces, at $z = \pm L$.  The corresponding roton wave function
is particularly simple at $k_x=0$,  where it takes the form
\begin{equation}
\Psi_r (z) \propto \cos(p_0 z) \cos ({\pi z \over 2 L}) \ \ \ .
\label{cos}
\end{equation}
This picture is well confirmed by the numerical solution of the
equations of motion. The comparison between the result of the
numerical calculation and formula (\ref{cos})
is shown in Fig.~\ref{cosinus}. The two curves
are almost indistinguishable, showing that the density
functional approach properly accounts  for the change-mode
reflection of rotons on the free surface.

We note however that result (\ref{ralmost1}), yielding change-mode
reflecton,  is not valid when  $\alpha(\omega,k_x)=0$. We can study
the behaviour of rotons also in this case using the arguments
of Sect. \ref{hybri}. In fact, for $\omega > \Delta$, the line
where $\alpha=0$ is fixed by the equation
\begin{equation}
{g^2 \over \epsilon_0(k_x) - \omega } = - b (k_x) \ \ \ .
\label{alfa0}
\end{equation}
It starts at the point $k_t$ and corresponds to the continuation
of the ripplon dispersion above $\Delta$ (this continuation is
clearly visible in our numerical results of Fig. \ref{allstates}).
This line is singular  for the reflection of rotons at the surface.
It is clear from Eqs.  (\ref{r1omega}) and (\ref{r2omega}) that,
when $\alpha=0$, one has
\begin{equation}
R_1 \simeq 0 \ \ \ \ , \ \ \ R_2 \simeq 1 \ \ \  .
\end{equation}
Thus the normal reflection dominates near the singular line.

We note again that in a slab of finite thickness $2L$ the roton levels
are discretized. For the lowest levels near $\Delta$  and far enough
from the singular line $\alpha=0$, the change-mode reflection is dominant.
The condition of energy quantization has the form
\begin{equation}
2 k_z L + 2 \delta = n \pi \ \ \ \ , \ \ \ n=1,2,3,\dots
\label{quantization}
\end{equation}
where $\delta$ is a phase shift at the reflection, defined by
\begin{equation}
R_1(\omega) = - |R_1| \exp[2i\delta] \ \ \ .
\label{defphase}
\end{equation}
In the region under consideration $R_1(\omega)\simeq -1$ and one can
choose $\delta=0$. Approaching the singular line $\alpha=0$ the
quantization condition  (\ref{quantization}) is no longer valid.
However, one can easily discuss the behaviour of the phase
during the crossing of this line. In fact, comparing Eq.
(\ref{defphase}) with Eq. (\ref{r1omega}), one can write
\begin{equation}
\tan (2\delta) =  - {k_z \over \alpha  } \ \ \ .
\end{equation}
Crossing the singular line from left to right $\alpha$ changes sign,
the change-mode reflection amplitude  is again equal to $-1$, but
the phase $\delta$ has increased by $\pi/2$. This means that the
orizontal line  $\omega_{(n)}(k_x)$ for a roton on the left
is matched to the line $\omega_{(n+1)}(k_x)$ on the right
of the singular line  $\alpha=0$.  These jumps are clearly seen in
Fig. \ref{allstates}, just above the point where the ripplon branch
reaches $\Delta$.

We finally note that
excited states in the region between the minimum roton energy and the
maxon energy have a finite amplitude outside the slab, resulting
from the coupling between rotons and evaporated atoms (see
Fig. \ref{threeexamples}). A detailed
analysis of this coupling can be done following the stabilization
method recently proposed in Ref.\ \cite{Man93}. One solves the eigenvalue
problem in a large box of variable size, containing the slab. The
energy of the excited states varies with the box size. From the analysis
of such energy dependence one can extract information about the width
of states localized in the slab, i.e., of rotons. A typical picture
is given in Fig.~\ref{bigboxsize}. The change of slope of each curve
corresponds to the coupling between free atoms and rotons,
related to the phenomenon of quantum evaporation. A detailed study
of this problem is in progress.

\section{ Conclusions}

In this work we have provided a detailed discussion of the ripplon
dispersion of superfluid \hef in the regime of short wave lengths. The
main results can be summarized as follows:

a) We have shown that the hybridization mechanism between ripplons
and rotons is dominated by the reflection properties of rotons from the
surface. We have discussed different scenarios for the behaviour
of the ripplon dispersion near the threshold $\omega=\Delta$, where
$\Delta$ is the usual roton gap.

b) We have calculated numerically the dispersion of ripplons
by solving the equations of time dependent density functional theory
in thick slabs. A phenomenological backflow term has been included in
the functional in order to provide a realistic dispersion of bulk
rotons. The resulting ripplon dispersion turns out to be in reasonable
agreement with experiments. Our calculations confirm in an explicit
way the general reflection mechanism exhibited by rotons.

An extension of the present analysis to investigate
higher energy processes associated with the phenomenon of quantum
evaporation will be the object of a future paper.

\acknowledgments
We are deeply indebted to C. Carraro for many useful discussions
and suggestions concerning the reflection properties of rotons from the
surface. Stimulating discussions with L. Pricaupenko and J. Treiner are
also acknowledged. A.L. likes to thank A. Bosin for useful computational
support.   L.P. likes to thank A. Larkin, H. Taylor and P. Wolfle for
stimulating discussions. He thanks the hospitality of the
Dipartimento di Fisica at the University  of  Trento as well as of
the Institute for Condensed Matter
Theory at the University of Karlsruhe. This work was
partially supported by the Humboldt Fundation.

\appendix
\section*{ }

In this appendix we provide a simple argument to derive
expression (\ref{alphabg}) for $\alpha(k_x,\omega)$. To this
purpose one must solve  the Schr\"odinger equation for the roton
wave function $f_i$ in the domain  $z \ll \kappa^{-1}$ including
the boundary. For simplicity, since the wave function $\Psi_r$
can be chosen real, we take $f=f_1=f_2^*$, so that
\begin{equation}
\alpha (\omega,k_x) = -  {f'(z) \over f(z)}\bigg\vert_{z=0} \ \ \ .
\label{alphaf}
\end{equation}
The Schr\"odinger equation for $f$ has the form
\begin{equation}
(H_f - \Delta) f =0 \ \ \ ,
\label{ahf}
\end{equation}
where the Hamiltonian $H_f(k_x)$ contains the  interaction with the surface.
Notice that in Eq. (\ref{ahf}) we have put $\omega=\Delta$ since the
$\omega$-dependence of the solution is smooth and can be ignored in
first approximation. Let the
value of $\alpha$ predicted by Eq. (\ref{ahf}) be $b(k_x)$.

The ripplon dispersion is defined by the Schr\"odinger equation for
the ripplon wave function $\phi$
\begin{equation}
(H_\phi - \omega) \phi = 0
\label{ahphi}
\end{equation}
whose solutions  define the unperturbed ripplon dispersion
$\epsilon_0(k_x)$.

Hybridization means that on the right hand side of Eqs. (\ref{ahf})
and (\ref{ahphi}) we should add  terms linear in $\phi$ and $f$, respectively:
\begin{equation}
(H_f -\Delta) f = \tilde{g} {\cal L} \phi
\label{ahf2}
\end{equation}
\begin{equation}
(H_\phi - \omega) \phi = \tilde{g} {\cal L}_1 f
\label{ahphi2}
\end{equation}
where ${\cal L}$ and ${\cal L}_1$ are some operators, while $\tilde{g}$
is a coupling constant. Clearly the last equation has a pole when
$\omega$ is equal to the eigenvalue $\epsilon_0(k_x)$ of $H_\phi$.
Near the pole  one can write the solution as
\begin{equation}
\phi \simeq  {\tilde{g} \over \epsilon_0 (k_x) - \omega } {\cal L}_2 f
\label{aphi}
\end{equation}
where the quantity ${\cal L}_2 f$ has no dependence on $\omega$. Equation
(\ref{ahf2}) then becomes
\begin{equation}
(H_f - \Delta) f =
{\tilde{g}^2 \over \epsilon_0(k_x) - \omega }
{\cal L} {\cal L}_2 f  \ \ \ .
\label{a6}
\end{equation}
The solution of Eq. (\ref{a6}), needed to calculate $\alpha(\omega,k_x)$,
is difficult in general.   However it is evident a priori that
the solution  depends on $\omega$ only through the combination
$\tilde{g}^2/(\epsilon_0(k_x)-\omega)$. One can finally calculate
$\alpha(\omega,k_x)$ by treating the
contribution  of the right hand side in Eq. (\ref{a6}) as a
perturbation. We then obtain the desired  result
\begin{equation}
\alpha(\omega,k_x) = b(k_x) + { g^2 \over \epsilon_0(k_x) -
\omega } \ \ \ ,
\label{basta}
\end{equation}
used in Sect. \ref{hybri}.

A non perturbative approach to the same result can be also formulated
using Green's function techniques.

%

%

\begin{figure}
\caption{Schematic picture of the change-mode reflection. A
roton with positive group velocity ($R^+$) reaches the free surface and
transforms in a roton with negative group velocity ($R^-$). }
\label{change-mode}
\end{figure}

\begin{figure}
\caption{Effective current-current interaction (adimensional) entering
functional  (40)}
\label{vj}
\end{figure}

\begin{figure}
\caption{Dispersion relation of bulk and free surface excitations.
The present result for the surface mode (lowest
solid  line) is compared with the
experimental data on films  [1] (squares) and with the
hydrodynamic dispersion of ripplons (dots). The bulk phonon-roton
branch (upper solid line)  is compared with the experimental one
(circles).  The threshold for roton states with different $k_x$
is shown as the horizontal line. The threshold for the emission  of
atoms into the vacuum is also shown (dot-dashed curve).}
\label{schematic}
\end{figure}

\begin{figure}
\caption{Same as in Fig.~2 but with the bulk and surface mode dispersions
calculated without the current-current interaction in the density
functional ($V_J \equiv 0$).}
\label{feynman}
\end{figure}

\begin{figure}
\caption{Dispersion relation of the excited states of a slab
$50$ \AA\ thick, as a function of the parallel wave vector.}
\label{allstates}
\end{figure}

\begin{figure}
\caption{Dynamic structure function (in arbitrary units) for $k_z=0$
and $k_x=0.8$ \AA$^{-1}$ in a slab $50$ \AA\ thick. The ripplon  mode
corresponds to the peak at lowest energy.}
\label{strength}
\end{figure}

\begin{figure}
\caption{Ratio between the strength of the surface mode and the total
strength at $k_x=0.8$ \AA$^{-1}$ as a function of $k_z$.}
\label{angle}
\end{figure}

\begin{figure}
\caption{Three examples of excited states of a slab $50$ \AA\ thick. The
dashed line is the static density profile. The solid line corresponds to
the numerical solution of the equations of motion, at given values of
energy and parallel wavevector. The three states have the same $k_x$ but
different energy: a) the lowest energy surface mode; b)  free atoms
outside the slab coupled to phonons; c) free atoms outside
the slab coupled to rotons.}
\label{threeexamples}
\end{figure}

\begin{figure}
\caption{Detail of the repulsion between the surface mode and the
bulk roton mode near the threshold energy $\Delta$, for a slab $50$ \AA\
thick. The form of the lowest eigenstate at two different values of
$k_x$ is also shown in the upper part of the figure.}
\label{gap}
\end{figure}

\begin{figure}
\caption{Wave function of the first excited state above the roton
threshold at $k_x=0$. It corresponds to a superposition of $R^+$ and
$R^-$ rotons which undergo a change-mode reflection at the surface.
Solid line: numerical solution of the equations of motion. Dashed
line: analytic expression (43). Short dashed line: static density profile.}
\label{cosinus}
\end{figure}

\begin{figure}
\caption{Energy of the excited states at $k_x=0.4$ \AA$^{-1}$ as a function of
the size of the computational  box, containing a slab of fixed thickness
($50$ \AA).  The almost horizontal part of each curve corresponds to
roton-like states inside  the slab; the part with negative slope
(proportional to $L_{\hbox{box}}^{-2}$) corresponds to free atom states
outside the slab. The form of the level repulsion is connected to the
probability for a roton to decay by ejecting a free atom from the liquid.
The energy of the surface modes (not shown in the figure) is practically
independent on $L_{\hbox{box}}$. }
\label{bigboxsize}
\end{figure}


\begin{references}
\bibitem{note2} Clearly Hamiltonian (\ref{hf}) is not applicable to
the point $k_x=p_0$, where $1/m^*$ vanishes. This case should be treated
separately.
\bibitem{note3} This result is valid in general when $k_z \to 0$ and
holds also when the rotons are damped by the process of quantum
evaporation of atoms. The damping
process can be taken into account by simply introducing an imaginary part
in $\alpha$. The limiting value $-1$ of the change-mode reflection amplitude
will be the same also in this case, so that the probability of evaporation
tends to zero for $\omega \simeq \Delta$.
\end{references}
\end{document}